# Thermal Management of Photovoltaics using Porous Nanochannels

Sajag Poudel, An Zou, Shalabh C. Maroo*

Department of Mechanical and Aerospace Engineering, Syracuse University, NY 13244

*correspondence: scmaroo@syr.edu

**Abstract**

The photoelectric conversion efficiency of a solar cell is dependent on its temperature. When the solar radiation is incident on the photovoltaics (PV) panel, a large portion of it is absorbed by the underlying material which increases its internal energy leading to the generation of heat. An overheated PV panel results in a decline in its performance which calls for an efficient cooling mechanism that can offer an optimum output of the electrical power. In the present numerical work, thermal management with a porous nanochannels device capable to dissipate high heat flux is employed to regulate the temperature of a commercial PV panel by integrating the device on the back face of the panel. The spatial and temporal variation of the PV surface temperature is obtained by solving the energy balance equation numerically. By evaluating the steady-state PV surface temperature with and without thermal management, the extent of cooling and the resulting enhancement in the electrical power output is studied in detail. The nanochannels device is found to reduce the PV surface temperature significantly with an average cooling of 31.5 $^{\circ}$C. Additionally, the enhancement in the electrical power output by ~33% and the reduction in the response time to 1/8$^{th}$ highlight the potential of using porous nanochannels as a thermal management device. Furthermore, the numerical method is used to develop a universal curve which can predict the extent of PV cooling for any generic thermal management device.

## 1. Introduction

Being a renewable source, solar energy is extensively utilized through a variety of conversion approaches including photoelectricity, solar thermal system, photosynthesis based solar fuel, biomass, etc. (Crabtree & Lewis, 2007). Of these, the photoelectrical conversion process is ought to be more advantageous regarding the ease of transmission and utilization, however achieving high efficiency in the very method is the most provocative aspect of the process. Unlike the solar thermal conversion process, which can be up to 60% efficient, the solar cell utilizes only a narrow range of wavelength (visible light) resulting to a typical conversion efficiency of < 20% in practical PV panels (Crabtree & Lewis, 2007) (or up to 35% in certain special circumstances (Ajuria et al., 2011; Essig et al., 2017). This adds a further challenge in reducing the gap that exists between the solar energy potential and the extent of utilization (Crabtree & Lewis, 2007). In order to enhance the performance of a PV panel, several methods have been accomplished including reducing the optical losses (Escarré et al., 2012; Forbes, 2012), employing a concentrated PV system (Lasich et al., 2009; Schuetz et al., 2012), minimizing the internal resistances of the solar cell (Al-Rifai, Carstensen, & Föll, 2002; Han et al., 2005), applying post-processing techniques (Ghimire et al., 2021; Tang et al., 2019; Jifeng Yuan, Zhang, Bi, Wang, & Tian, 2018), using organic semiconductor (Cheng, Yang, & Hsu, 2009; Xue, Zhang, Li, & Li, 2018; Jun Yuan et al., 2019) and low bandgap materials (Boudreault, Najari, & Leclerc, 2011; Dou et al., 2012), etc. Besides, cooling of the PV panel is projected to be the best alternative especially while considering the establishment of the PV plants in tropical zones with elevated ambient temperature (Daher, Gaillard, Amara, & Ménézo, 2018; Dewi, Risma, & Oktarina, 2019; Ogbomo, Amalu, Ekere, & Olagbegi, 2017), which unsurprisingly implies overheating of the PV panels.

When the solar radiation is incident on the solar cells in a photovoltaic (PV) module, only a small part of the absorbed energy is transformed into electrical power while a significant part raises the internal energy of the underlying material. The rise in internal energy generates heat, which reduces the overall efficiency of the photoelectric conversion process (D. Du, Darkwa, & Kokogiannakis, 2013; Hasanuzzaman, Malek, Islam, Pandey, & Rahim, 2016) as well as diminishes the operating life (Sharma, Gupta, Nandan, Dwivedi,

& Kumar, 2018) of the PV panel. The conversion efficiency of a solar cell can vary widely with 3 – 20 % of the incident solar radiation transformed into electrical power (Crabtree & Lewis, 2007; Green et al., 2018). Based on the efficiency, a huge portion of the inbound energy gets dissipated in the form of heat and can raise the PV panel temperature up to 75 °C (Abd-Elhady, Serag, & Kandil, 2018). Several studies have highlighted overheating as the major hindrance to achieving a high conversion efficiency in a PV panel (Abd-Elhady et al., 2018; Bahaidarah, Baloch, & Gandhidasan, 2016; D. Du et al., 2013; Najafi & Woodbury, 2013; Waqas & Ji, 2017). Additionally, findings have reported the decline in the conversion efficiency by 0.5% - 0.7% per 1 °C rise in the temperature (Abd-Elhady et al., 2018; Rabie, Emam, Ookawara, & Ahmed, 2019) which further indicate a pressing need for an efficient cooling mechanism to regulate the PV panel temperature. Moreover, the temperature regulation of the PV surface potentially prevents any mechanical failure of the solar cell material due to thermal stresses as well (Bahaidarah et al., 2016). Finally, the outcome of the applied cooling in the PV panel in terms of the solar cell performance acts cumulatively to the other methods (Rejeb et al., 2020; Vaillon, Dupré, Cal, & Calaf, 2018; Yan, Ye, & Seferos, 2018), which further emphasizes the significance of exploring the PV cooling system. Cooling mechanisms can also be made financially viable by tuning the coolant supply (Castanheira, Fernandes, & Branco, 2018; Hadipour, Zargarabadi, & Rashidi, 2021) or utilizing a passive flow (Mittelman, Alshare, & Davidson, 2009; Nižetić, Papadopoulos, & Giama, 2017).

Several studies (Mazón-Hernández, García-Cascales, Vera-García, Káiser, & Zamora, 2013; Mittelman et al., 2009; Najafi & Woodbury, 2013) have shown efficient cooling of the PV panel through an air-cooling system demonstrating a lesser need of energy to circulate the air coolant than liquid. However, a higher extent of cooling can be achieved with water cooling (Bahaidarah et al., 2016; Castanheira et al., 2018; Jakhar, Soni, & Gakkhar, 2017; Nižetić et al., 2017) especially methods employing phase change (Alami, 2014; Chandrasekar & Senthilkumar, 2015; Nižetić, Giama, & Papadopoulos, 2018) that removes a very large amount of latent heat. (Jakhar et al., 2017) employed water channels at the back face of the PV panel to continuously circulate liquid coolant. Similarly, jet impingement on the PV panel was introduced, which was capable to reduce the PV surface temperature from 69.7 °C to 36.6 °C (Bahaidarah, 2016). With an aim to reduce the consumption of coolant, a periodic cycle of cooling by switching the sprinklers on/off was applied, which reported an enhancement in the annual energy production by 12% (Castanheira et al., 2018). Additionally, a system of pulsed-spray cooling also showed an effective enhancement of 27% in the electrical power output by consuming only 1/9$^{th}$ of the coolant that would be used in a steady supply (Hadipour et al., 2021). In order to achieve higher cooling proficiency by exploiting latent heat, various phase change materials have also been utilized to regulate the PV panel temperature demonstrating an average cooling of 20 – 30 °C (Huang, Eames, & Norton, 2006; Waqas & Ji, 2017). Furthermore, a passive dissipation of heat through thin-film evaporation was achieved by using layers of synthetic clay (Alami, 2014) as well as cotton wick structures (Chandrasekar & Senthilkumar, 2015) on the back face of the PV panel ultimately enhancing the electrical power output by 19% and 14% respectively. Unlike the active supply of coolant, the passive flow of liquid by employing wick structures (Alami, 2014; Chandrasekar & Senthilkumar, 2015) or heat pipes (Alizadeh et al., 2018; Shittu et al., 2019), etc. do not need supplemental power to drive the flow of coolant. While the power consumed to drive the coolant in an active cooling system can consume a substantial portion of the generated electricity and reducing the enhancement in the electrical power by ~50% (Bai et al., 2016; Nižetić, Čoko, Yadav, & Grubišić-Čabo, 2016), passive cooling system eliminates the burden to drive such external flow. With this understanding, the passive supply of liquid and heat flux dissipation through phase change are recognized as two key criteria for an efficient PV cooling system. In line with that, a device with micro/nano scale structures that can (1) continuously supply required liquid coolant through wicking (Krishnan, Bal, & Putnam, 2019; Lee, Suh, Dubey, Barako, & Won, 2018; Poudel, Zou, & Maroo, 2019) and (2) simultaneously allow a high heat removal rate at the evaporating menisci (Fischer et al., 2017; Jasvanth, Ambirajan, Adoni, & Arakeri, 2019; Poudel, Zou, & Maroo, 2020b) is anticipated as the best candidate to offer a promising solution to the overheating issue of the PV panel. Accordingly, we utilize the porous nanochannels device (Poudel, Zou, & Maroo, 2020a) which has demonstrated excellent wicking characteristic (Poudel et al., 2019, 2020b) as well as high heat flux dissipation through nanoscale thin-film evaporation (Poudel et al., 2020a), and numerically integrate it on the back face of a commercial PV panel to evaluate the extent of cooling. The cross-connected geometry of the buried nanochannels (height = 728 nm) offer a potential solution for high rate of wicking while the micropores (diameter = 2.1 μm) provided at each intersection host the sites for evaporating menisci (Poudel et al., 2019, 2020a; Zou, Poudel, Raut, & Maroo, 2019). Hence, the findings of the heat flux removal attained in such porous nanochannels device (Poudel et al., 2020a) is utilized as a technique of thermal management and the numerical investigation of PV cooling employing energy balance model (Y. Du et al., 2016) is reported here.

## 2. Methods

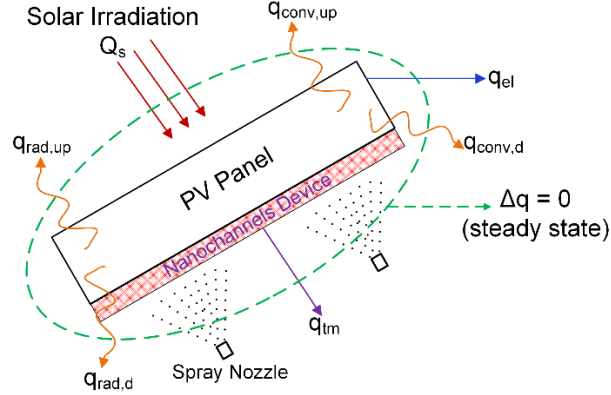

Figure 1. Schematic of a PV panel with nanochannels device attached on the back face illustrating the associated quantities of heat and energy transfer.

The energy balance model (Y. Du et al., 2016) is utilized to numerically solve the governing equations of energy and heat transfer in order to determine the temperature distribution over the surface of the PV panel. When the incoming solar radiation flux ($Q_s$) is incident on the PV panel (Fig. 1), a part of it will cause the rise in internal energy of the PV panel material while some portion would be lost to the ambient via convection ($q_{conv}$) and radiation ($q_{rad}$) heat transfer per unit area. A fraction of the incoming radiation flux is converted into electrical power per unit area ($q_{el}$). Accordingly, the energy balance model is expressed as shown in Eq. 1.

$$q_s - q_{el} - q_{conv} - q_{rad} - q_{tm} = \rho C_p \delta \frac{dT_s}{dt} \quad \text{Equation 1}$$

where, each term has the unit of W/m², $q_s = \varepsilon_o Q_s$, $\varepsilon_o = 0.9$ is the PV surface absorptivity, δ is the thickness of PV panel, and $q_{el}$ is the electrical power output from the PV panel per unit area which is dependent on the conversion efficiency of the PV panel ($\beta$) and is the function of the surface temperature $T_s$ (Alizadeh et al., 2018; Y. Du et al., 2016):

$$q_{el} = \beta Q_s = (21.737 - 0.1757 T_s) Q_s \quad \text{Equation 2}$$

$q_{conv}$ is calculated as the sum of convection heat transfer at the front and back face of the PV panel:

$$q_{conv} = q_{conv,up} + q_{conv,d} \sim 2 q_{conv,up} = 2h(T_s - T_a) \quad \text{Equation 3}$$

where, $T_s$ is the average surface temperature of the PV panel and $T_a$ is the ambient temperature ($T_a$ = 18ºC), $h$ is the convection heat transfer coefficient, which is the function of wind speed ($u_w$) as $h = 2.8 + 3.8 u_w$ and $u_w$ = 2 m/s (Alizadeh et al., 2018; Y. Du et al., 2016).

Similarly,

$$q_{rad} = q_{rad,up} + q_{rad,d} \cong 2q_{rad,up} = 2\varepsilon_l \sigma(T_s^4 - T_a^4) \qquad \text{Equation 4}$$

where, $\varepsilon_l$ represents the emissivity of the PV panel surface ($\varepsilon_l \sim 0.7$) (Alizadeh et al., 2018; Y. Du et al., 2016)

$q_{tm}$ is the heat flux removal achieved through thermal management, which is obtained by employing the porous nanochannels device (Poudel et al., 2020a) (henceforth denoted as nanochannels). The nanochannels utilized in the present study offer a passive dissipation of high heat flux through thin-film evaporation of the spray droplets dispersed over it, thus eliminates the need of a continuous supply of coolant. This further prevents the loss of energy in driving the active flow and simplifies the analysis relating the economic viability of the energy management system.

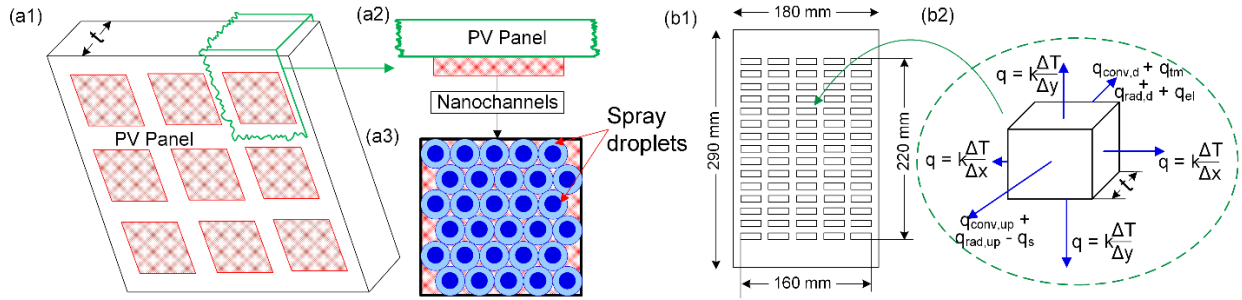

Figure 2. Detail of PV cooling system with nanochannels. (a1-a2) Integration of nanochannels devices on the back face of the PV Panel. (a3) Spray droplets dispersed over the nanochannels device. (Poudel et al., 2020a) (b1) The commercial PV panel utilized in present study. (Alizadeh et al., 2018) (b2) A unit control volume in the discretized domain of the PV panel illustrating the associated transport quantities.

In order to provide a suitable thermal management, the nanochannels are integrated on the back face of the PV panel to dissipate heat from the material of the PV panel. In order to achieve uniformity in the surface temperature, several such nanochannels are attached to the PV panel (see Fig. 2-a1), on top of which, the dispersed spray droplets wick and evaporate passively (see Figs. 2-a2 and 2-a3). Moreover, the number and distribution of the nanochannels device attached to the PV panel can be varied to tune the required extent of cooling. Figure 2-b1 shows the geometry of a commercial PV panel (size of 180 mm × 290 mm and thickness of 5.45 mm) employed for the numerical study in the present work. The PV panel having the monocrystalline silicon solar cells with same geometry has been utilized in several other studies (Alizadeh et al., 2018; Y. Du et al., 2016; García & Balenzategui, 2004). Information about the material, fabrication and performance of the solar cells in this PV panel is adapted from literature (Alizadeh et al., 2018; Y. Du et al., 2016).

Next, to solve the energy balance equation (Eq. 1), the domain of the PV panel (Fig. 2-b1) is discretized into a finite number (N) of control volumes. For each control volume, the energy balance equation (Eq. 1) is deduced with an additional term to account for the conduction heat transfer ($q_{cond}$) along the length and width of the PV panel as illustrated in Eq. 5. Fig. 2(b2) demonstrates all transport quantities associated with a typical inner control volume. For the control volume at the edge of the PV panel, boundary condition applied specified is convection + radiation. Additionally, the initial temperature of the entire domain is set as uniform and equal to the ambient temperature ($T_a$ = 18 °C) and the properties of the PV panel material is assumed to be independent of temperature in the operating range.

$$q_s - q_{el} - q_{conv} - q_{cond} - q_{rad} - q_{tm} = \rho C_p \delta \frac{dT_s}{dt} \qquad \text{Equation 5}$$

By solving the transient state energy balance equation (Eq. 5), the surface temperature at the center of each control volume of the discretized domain is achieved through numerical iterations. The temperature distribution is further used to characterize the average surface temperature of the PV panel ($T_s$) at corresponding time instant ($t$). Initially, by varying the number of control volumes, multiple cases of numerical simulation with $Q_s$ = 1,000 W/m² is studied and the evolution of $T_s$ with time is compared to investigate the grid-sensitivity. From the grid sensitivity test, a grid with 200×400 (N = 80,000) control volumes is identified as the optimum one, which is then employed for all subsequent cases.

## 3. Results and Discussion

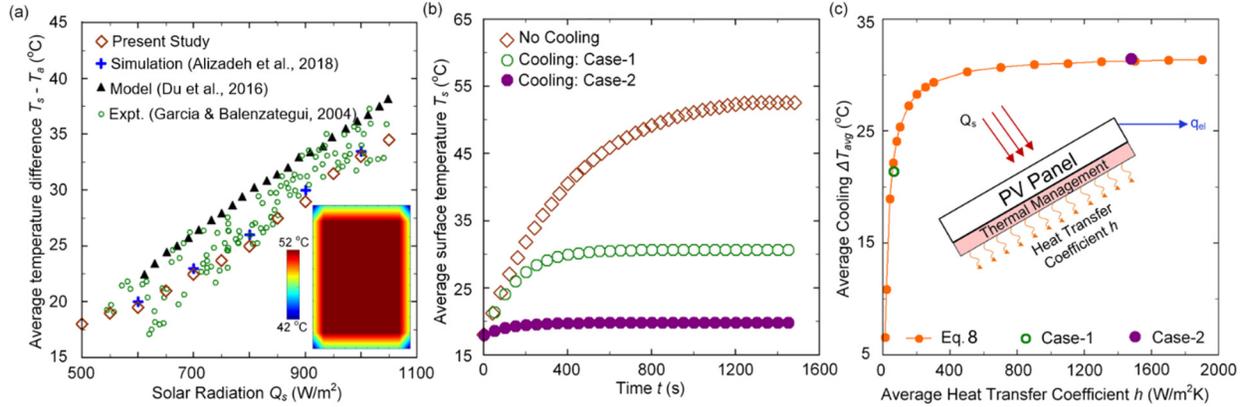

Figure 3. Thermal state of the reference PV panel at varying conditions. (a) Variation of the difference in average surface temperature and ambient temperature with solar radiation. The inset shows the steady state temperature distribution over the PV panel with $Q_s$ = 1,000 W/m². (b) Temporal rise in average surface temperature of the PV panel for the cases of cooling and no cooling. (c) Variation of the average cooling achieved with the heat transfer coefficient for a generic thermal management.

Initially, the energy balance equation (Eq. 5) is solved for the temperature distribution on the PV panel by varying $Q_s$ = 500 W/m² to 1,050 W/m² and without any thermal management ($q_{tm} = 0$). As the PV panel temperature reaches a steady state, the difference in average temperature of the PV panel surface ($T_s$) and the ambient temperature ($T_a$) is obtained for each case of solar radiation and plotted in Fig. 3a. The inset in Fig. 3a also depicts the steady state temperature distribution over the PV panel surface corresponding to $T_s$ = 1,000 W/m² at $t$ = 1,500 s. The obtained variation of the average temperature difference ($T_s - T_a$) with the incident solar radiation shows excellent agreement with the results of experiments (García & Balenzategui, 2004), numerical simulation (Alizadeh et al., 2018) and analytical model (Y. Du et al., 2016) on the same PV panel as shown in Fig. 3a. Thus, the plot in Fig. 3a establishes the accuracy and reliability of the numerical methods implemented in the present study.

When the specified solar radiation is incident to the PV panel, temperature of the PV panel surface ($T_s$) escalates from initial 18 °C due to the rise in internal energy of the PV material. The surface temperature ultimately approaches a steady state in a certain duration of time designated as response time. Response time ($R_t$) is demarcated as the time duration at which the PV panel surface temperature reaches 99% of the final steady state temperature (Y. Du et al., 2016) due to the combined effect of gain in internal energy (owing to incoming radiation) and loss in heat (due to thermal management). Accordingly, Fig. 3b shows the temporal evolution of $T_s$ corresponding to $Q_s$ = 1,000 W/m² with the final steady state temperature of the PV panel being ~51.5°C with $R_t$ ~ 1,500 s.

Next, the heat flux dissipation due to the employed nanochannels is introduced for the thermal management of the PV panel. The performance of the nanochannels in two different cases of spray cooling corresponding to the uniform spray droplets diameter 400 μm and 20 μm (denoted by Case-1 and Case-2 respectively) achieved earlier (Poudel et al., 2020a) is utilized here. The amount of heat flux removal during such process is found to be dependent on the surface temperature; Accordingly, following two expression for the heat flux as a function of $T_s$ is deduced from the past work (Poudel et al., 2020a):

$$\log_{10}(q_{tm}) = 0.03369T_s - 2.143 \qquad \text{Equation 6}$$

$$\log_{10}(q_{tm}) = 0.03036T_s - 1.577 \qquad \text{Equation 7}$$

The nanochannels are incorporated to the entire surface of the back face of the reference PV panel such that maximum and uniform heat flux dissipation is possible. When this applied thermal management ($q_{tm}$) is introduced to the energy balance equation (Eq. 5), the surface temperature of the PV panel reduces. The variation of $T_s$ with time for the two cases of cooling (dictated by Eqs. 6 and 7) with $Q_s$ = 1,000 W/m² is also shown in Fig. 3b. The plot demonstrates the final steady state temperature to be to 30.5°C and 20°C for Case 1 and 2 respectively as compared to 51.5°C for the case without any cooling. In addition, as compared to $R_t$ ~ 1,500 s for no cooling case, the system with cooling, cases-1 and 2 reach the steady state of operation much sooner with $R_t$ ~ 500 s and ~ 200 s respectively. It is essential to have a shorter $R_t$ as the incoming radiation varies throughout the day and the thermal hysteresis are not trivial. A shorter response time achieved with the implemented thermal management further signifies the impact of the results reported here.

Next, we utilize the energy balance equation (Eq. 5) to compute the PV surface temperature distribution by considering a case of thermal management with specified heat transfer coefficient ($h_{tm}$). While the investigation of PV cooling by employing nanochannels provide the extent of cooling accomplished through corresponding removal of heat flux (Eqs. 6 & 7), the consideration of $h_{tm}$ can provide an estimate of the cooling for a broader range of heat transfer concerning generic applications. The implication of thermal management with specified $h_{tm}$ on PV cooling is crucial to comprehend the potential cooling performance of widespread techniques including phase change material, nanofluids, wickless heat pipes, etc. (Verma, Mohapatra, Chowdhury, & Dwivedi, 2021). In order to resolve this, numerical study of PV cooling for a range of $h_{tm}$ = 1 – 2,000 W/m²K is performed by characterizing $q_{tm}$ as:

$$q_{tm} = h_{tm}(T_s - T_a) \qquad \text{Equation 8}$$

Figure 3c shows the variation of the average cooling achieved with the heat transfer coefficient. Here, the extent of cooling achieved is characterized by average cooling ($\Delta T_{avg}$) such that $\Delta T_{avg} = T_{s-wo} - T_{s-w}$, where $T_{s-w}$ and $T_{s-wo}$ are the steady state average surface temperature of the PV panel with and without thermal management respectively. This plot signifies a wide-ranging implication and can be utilized to estimate the average cooling of the PV panel obtained via a generic thermal management device of known heat transfer coefficient ($h$). Furthermore, the plot in Fig. 3c exhibits a unique nature of variation with $h_{tm}$; the initial rise in $h_{tm}$ shows a steep increment in $\Delta T_{avg}$ but ultimately approaches its maximum limit. In such, a system will attain the maximum extent of cooling where a further increment in $h_{tm}$ doesn't enhance $\Delta T_{avg}$ anymore, hence inferring an upper limit of the extent of the PV panel cooling. This is an exciting result obtained in regard that it predicts the theoretical limit of the PV panel cooling which is possible to establish only by characterizing $q_{tm}$ in terms of $h_{tm}$ in Eq. 5. Additionally, considering the steady state operation of the two cases of the PV panel cooling corresponding to $Q_s$ = 1,000 W/m² (see. Fig. 3b), we back calculate the heat transfer coefficient from (Eqs. 6, 7) as $h_{tm}$ = $q_{tm}/(T_s - T_a)$. The average cooling versus $h_{tm}$ for these two cases are also presented in the same plot Fig. 3c. Clearly, the finer spray droplets in Case-2 provides a higher heat flux dissipating ability (Eq. 7) and exhibits a greater extent of PV panel cooling approaching the theoretical limit. Overall, Fig. 3c provides a universal standard to estimate PV cooling for a generic thermal management device and indicates the cooling ability of the employed nanochannels.

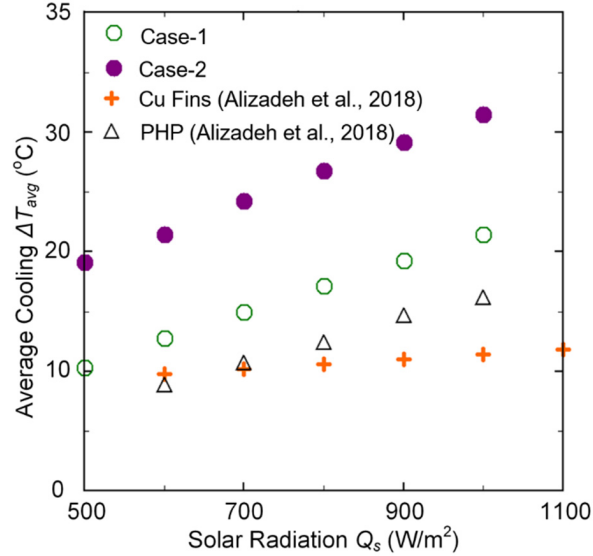

Figure 4. Variation of average cooling of the PV panel with incident solar radiation.

Figures 3b and 3c provide the results of PV cooling for a constant value of solar radiation ($Q_s$ = 1,000 W/m$^2$). However, in a realistic scenario, the intensity of solar radiation fluctuates with weather and is time dependent. Thus, it necessitates the investigation of the average cooling $\Delta T_{avg}$ for a varying range of $Q_s$. Considering the solar radiation variation $Q_s$ = 500 W/m$^2$ to 1,000 W/m$^2$ with step of 100 W/m$^2$, the thermal management with nanochannels (Cases 1 and 2) is employed over again to solve Eq. 5 numerically. Figure 4 shows the variation of average cooling accomplished with the incident solar radiation. A significant extent of cooling of the PV panel using the nanochannels in the present study is achieved throughout the range of solar radiation considered. Moreover, Fig. 4 also demonstrates the superior cooling with the present method as compared to the past work of PV cooling achieved through Cu Fins and pulsating heat pipe (PHP) (Alizadeh et al., 2018) on the same PV panel. Such an improvement in the present work is possible due to thin-film evaporation based phase-change heat transfer occurring in the porous nanochannels device (Poudel et al., 2020a). Further, the enhancement in Case-2 as compared to Case-1 also aligns well with the finding in terms of the heat transfer coefficient (see Fig. 3(c)) thus reiterating the significance of the reported technique of PV panel cooling.

Following the findings of the PV panel cooling with the nanochannels, we report the performance of the PV panel in terms of electrical power output. As the employed thermal management reduces the surface temperature of the PV panel, this in turn enhances the photoelectric conversion efficiency and the electrical power output of the system. By considering the steady state average temperature of the PV panel surface ($T_s$), Eq. 2 is utilized to compute the electrical power output for the range of solar radiation considered in the present study. The chart in Fig. 5 shows the electrical power output attained with the cooling via nanochannels (Cases 1 and 2) as compared to the case without any thermal management. The chart also shows the numerically obtained results for the steady state electrical power output of the consistent PV panel without employing any cooling as reported in a recent study (Alizadeh et al., 2018). In the range of $Q_s$ considered, average enhancement of $\varepsilon_{el}$ = 32.8% in the electrical power output is achieved with the employed thermal management corresponding to Case-2. The reported $\varepsilon_{el}$ in the present work surpasses that in many other techniques of PV panel cooling including active spray ($\varepsilon_{el}$ ~7%) (Nižetić et al., 2016), pulsating heat pipes ($\varepsilon_{el}$ ~18%) (Alizadeh et al., 2018), phase change material ($\varepsilon_{el}$ ~23%) (Hasan, Sarwar, Alnoman, & Abdelbaqi, 2017), pulsed water spray ($\varepsilon_{el}$ ~27%) (Hadipour et al., 2021), hybrid photovoltaics-thermoelectric-heat system ($\varepsilon_{el}$ ~30%) (Yang & Yin, 2011), etc.

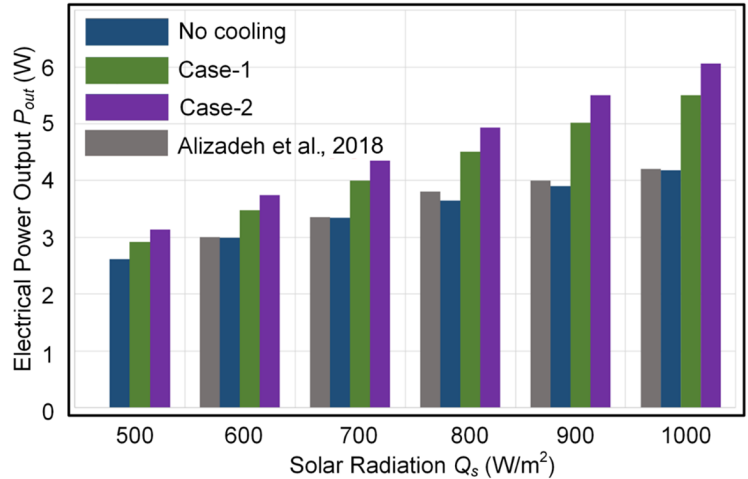

Figure 5. Electrical power output of the PV panel for various scenarios of thermal management at corresponding solar radiation.

## 4. Conclusion

This paper presents a numerical investigation of photovoltaics (PV) panel cooling by employing spray-cooling heat flux dissipation on porous nanochannels integrated on the back face of the PV panel. The energy balance equation on the PV panel system is solved numerically to obtain the spatial and temporal variation of the temperature over the PV surface for different cases with and without thermal management. The extent of cooling achieved by applying thermal management with nanochannels device for a range of incident solar radiation is studied in detail. The numerical method is also used to deduce a universal curve which can predict the magnitude of PV cooling of any generic thermal management technique. The universal curve revealed the existence of a theoretical limit for the extent of cooling. Finally, the employed nanochannels device exhibited an excellent performance with an average cooling of up to 31°C indicating its performance to be near the predicted theoretical limit. The overall enhancement in the electrical power output by 32.8% is obtained which surpasses other similar cooling methodologies reported in the literature.

**Acknowledgment**

The study presented in this paper is based on the work supported by the Office of Naval Research under contract/grant no. N000141812357.